\documentclass{hep99}
\usepackage{epsfig}
\begin{document}
\title{New results on interference effects and correlations}
\author{E J Thomson}
\address{Physics Department, 
Imperial College of Science, Technology and Medicine,\\
Prince Consort Road, LONDON SW7 2AZ, UK\\[3pt]
E-mail: {\tt evelyn.thomson@cern.ch}}

\abstract{New results on interference effects and correlations are
reviewed, including Fermi-Dirac and Bose-Einstein correlations in
hadronic $Z^{0}$\ decays at LEP1.} 
\maketitle

\section{Introduction}

Studies of Bose-Einstein (BE) correlations of identical bosons and of
Fermi-Dirac (FD) correlations of identical fermions produced in high
energy collisions provide measurements of the distributions of
particle sources in space and time.  These correlations originate from
the symmetrization or antisymmetrization of the two-particle wave
functions of identical particles and lead to an enhancement or a
suppression of particle pairs produced close to each other in phase
space.  The strength of two-particle BE or FD correlation effects can
be expressed in terms of a two-particle correlation function
$C(p_{1},p_{2})$\ defined as:

\begin{displaymath}
C(p_{1},p_{2}) = P(p_{1},p_{2}) / P_{o}(p_{1},p_{2})
\end{displaymath}

\noindent where $p_{1}$\ and $p_{2}$\ are the four-momenta of the
particles, $P(p_{1},p_{2})$\ is the measured differential
cross-section for the pairs and $P_{o}(p_{1},p_{2})$\ is that of a
reference sample, which is free of BE or FD correlations but otherwise
identical in all aspects to the data sample.  The main experimental
difficulty is to define an appropriate reference sample
$P_{o}(p_{1},p_{2})$\ in order to determine that part of 
$P(p_{1},p_{2})$\ which can be attributed to the BE or FD
correlations.  The correlation function $C$\ is usually measured as a
function of the Lorentz invariant $Q$, with $Q^{2} =
-(p_{1}-p_{2})^{2}$.  For $Q^{2}$ = 0, the effects of BE or FD
correlations reach their extreme values.

	Section 2 describes a new result on FD correlations
in $\Lambda\Lambda$\ and $\bar{\Lambda}\bar{\Lambda}$\ pairs.  
Section 3 discusses new results on multi-dimensional 
BE correlations in $\pi^{+}\pi^{+}$\ and $\pi^{-}\pi^{-}$\
pairs.  Section 4 summarises other new results and 
section 5 presents conclusions.

\section{FD correlations in $\Lambda\Lambda$\ 
and $\bar{\Lambda}\bar{\Lambda}$\ pairs}

\begin{figure}[!b]
\begin{center}
\mbox{\epsfig{figure=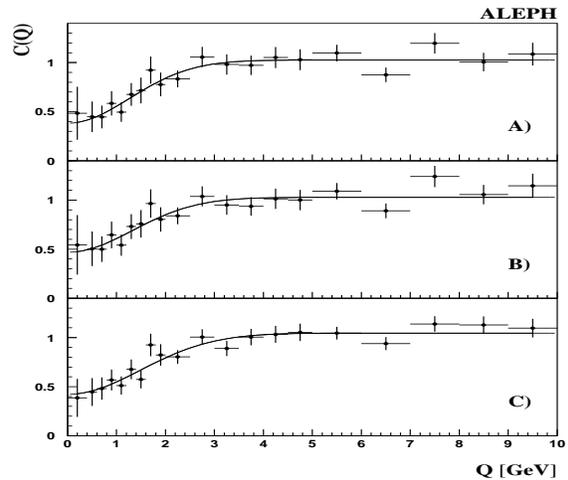,width=0.47\textwidth,height=0.3\textheight}}
\end{center}
\caption{\label{fig:fd3ref} Correlation function $C(Q)$\ for the
($\Lambda\Lambda$,$\bar{\Lambda}\bar{\Lambda}$) pairs using different
reference samples: A) Monte Carlo, B) mixed event double ratio and C)
mixed events with reweighted $\cos\theta_{1,2}$\ distribution. The
curves represent the results of fits using the Goldhaber parameterisation~\cite{alephfd}.}
\end{figure}

\begin{figure}[!b]
\begin{center}
\mbox{\epsfig{figure=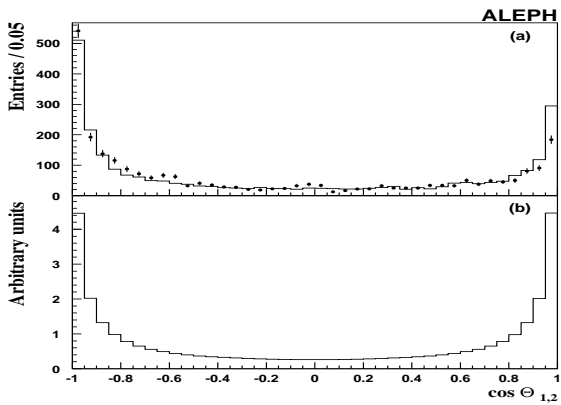,width=0.47\textwidth,height=0.25\textheight}}
\end{center}
\caption{\label{fig:fdcostheta} The histogram shows the
$\cos\theta_{1,2}$\ distribution for a) the original 
($\Lambda\Lambda$,$\bar{\Lambda}\bar{\Lambda}$) pairs from the Monte
Carlo and b) for the mixed pairs of the same sample.  Points in a)
show the $\cos\theta_{1,2}$\ distribution of the data.  The deviation
of the Monte Carlo from data at large $\cos\theta_{1,2}$\ is due to
the FD effect.
}
\end{figure}

ALEPH have measured the two-particle correlation function of
($\Lambda\Lambda$,$\bar{\Lambda}\bar{\Lambda}$) pairs~\cite{alephfd},
using three different references as shown in figure~\ref{fig:fd3ref}.  
All three references rely to some extent 
on a proper description of the hadronization
process by Monte Carlo.  Reference A
assumes the differential cross-section of simulated
($\Lambda\Lambda$,$\bar{\Lambda}\bar{\Lambda}$) in the JETSET Monte
Carlo.  References B and C are obtained by the technique of event mixing,
where pairs of $\Lambda$'s or $\bar{\Lambda}$'s are constructed by
pairing each $\Lambda$\ or $\bar{\Lambda}$\ with the $\Lambda$'s or
$\bar{\Lambda}$'s of all other events.  However, this method removes
not only possible FD or BE correlations, but it also affects all other
correlations as can be seen from figure~\ref{fig:fdcostheta}, where
the cosine of the angle $\theta_{1,2}$\ between the $\Lambda$\ or
$\bar{\Lambda}$\ momenta in the $Z^{0}$\ rest frame is plotted.  In
the unmixed events most pairs are produced back to back, whereas the
distribution for the mixed sample is completely symmetric.  This leads
to a shift in the $Q$\ distribution for the mixed pairs towards lower
values of $Q$.  To overcome this problem, the double ratio of the
cross sections is commonly used and this gives reference B:

\begin{displaymath}
C(Q) = \left(\frac{P(Q)_{\rm data}}{P(Q)_{\rm data,mix}}\right) \left/
\left(\frac{P(Q)_{\rm MC}}{P(Q)_{\rm MC,mix}}\right) \right.
\end{displaymath}

\noindent Since the symmetrization of the $\cos\theta_{1,2}$\
distribution is the main reason for the difference in the Q
distributions of the mixed and original pairs, reference C is
constructed from the mixed data sample, which is reweighted to
reproduce the $\cos\theta_{1,2}$\ distribution from the original pairs
obtained from Monte Carlo.  

Independent of the reference sample used, 
$C(Q)$\ shows a decrease for $Q<2$~GeV.
If this is interpreted as a FD effect, the size of the source $R$\ 
estimated from
$C(Q)$\ with a Goldhaber parameterisation
is:

\begin{eqnarray*}
R(\Lambda\Lambda,\bar{\Lambda}\bar{\Lambda}) & 
= & 0.11 \pm 0.02_{stat} \pm 0.01_{syst}\ {\rm fm}\\
\end{eqnarray*}

This is consistent with the result of the 
less precise but less model dependent measurement of the 
spin composition~\cite{alexlipkin} of the
($\Lambda\Lambda$,$\bar{\Lambda}\bar{\Lambda}$) system, which can have
a total spin of 0 or 1. 
The spin 1 fraction, $\epsilon(Q)$, is expected to be 0.75 for a
statistical spin mixture ensemble.  Since the total wave
function for the fermion pair must be antisymmetric, the symmetric spin
1 wave function must be paired with the antisymmetric space wave
function, and thus $\epsilon(Q)$ is expected to decrease to 0 as $Q$
goes to 0 due to FD statistics. 
ALEPH finds that $\epsilon<0.75$\ for $Q<2$~GeV and $\epsilon\simeq0.75$\ for
$Q>2$~GeV, as shown in Table~\ref{tab:fdepsi}.  The size of
the source $R$\ estimated from $\epsilon(Q)$\ is 
$R=0.14 \pm 0.09_{stat} \pm 0.03_{syst}$~fm.  For the
$\Lambda\bar{\Lambda}$\ system, which is free of FD correlations, the
spin composition measurements are consistent with $\epsilon$ = 0.75 in
the entire $Q$\ range studied.  ALEPH also proves that this spin
composition technique, previously shown to hold approximately for low
values of Q, is in fact valid for any value of Q.  The results for
$\epsilon(Q)$\ are in agreement with previous measurements from
OPAL~\cite{ofd} and DELPHI~\cite{dfd}.

\begin{table}[!b]
\begin{center}
\caption{\label{tab:fdepsi} The values of $\epsilon$\ for the 
($\Lambda\Lambda$,$\bar{\Lambda}\bar{\Lambda}$) and
$\Lambda\bar{\Lambda}$\ samples.}
\begin{footnotesize}
\begin{tabular}{ccc} 
\br
$Q$ Range [GeV] &
$\epsilon$($\Lambda\Lambda$,$\bar{\Lambda}\bar{\Lambda}$) &
$\epsilon$($\Lambda\bar{\Lambda}$) \\
\mr
0.0 - 1.5 & 	$0.36 \pm 0.30 \pm 0.08$ & 
		$0.61 \pm 0.13 \pm 0.07$ \\

1.5 - 2.0 &  	$0.52 \pm 0.31 \pm 0.10$ & 
		$0.77 \pm 0.07 \pm 0.03$ \\

2.0 - 4.0 & 	$0.78 \pm 0.16 \pm 0.09$ & 
		$0.51 \pm 0.11 \pm 0.12$ \\
\br
\end{tabular}
\end{footnotesize}
\end{center}
\end{table}

A comparison of this result 
with the measured radii for identical charged pions~\cite{afdpi} 
and kaons~\cite{afdk}:

\begin{eqnarray*}
R(\pi^{\pm},\pi^{\pm}) & = & 0.65 \pm 0.04_{stat} \pm 0.16_{syst}\ \rm{fm} \\
R(K^{\pm},K^{\pm})     & = & 0.48 \pm 0.04_{stat} \pm 0.07_{syst}\ \rm{fm} \\
\end{eqnarray*}

\noindent indicates that the dimension of the source decreases with
the increasing mass of the emitted particles.

\section{Multi-dimensional BE correlations}

A recent model of BE correlations predicts that the
longitudinal correlation length is different from the transverse one
since momentum components longitudinal and transverse with respect to
the string direction are generated by different mechanisms~\cite{andersson}.
To verify this prediction, 
OPAL~\cite{opalbe}, DELPHI~\cite{delphibe} and L3~\cite{l3be} have all investigated multi-dimensional
BE correlations in hadronic $Z^{0}$\ decays in the
so-called Longitudinally CoMoving System,
which represents the local rest frame of the string. 

\begin{figure}[!b]
\begin{center}
\mbox{\epsfig{figure=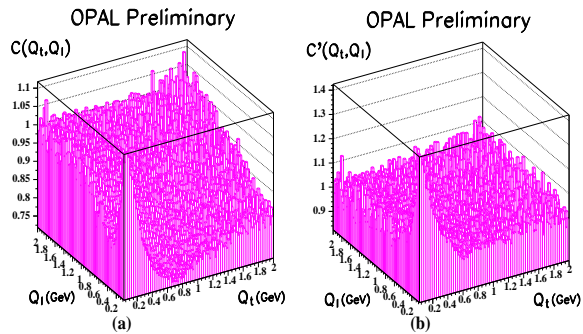,width=0.49\textwidth,height=0.2\textheight}}
\end{center}
\caption{\label{fig:opalbe} Two-dimensional BE
correlation functions plotted versus the two components of the momentum
difference, along the thrust axis, $Q_{l}$, and transverse to the same
axis, $Q_{t}$.}
\end{figure}

OPAL use a sample of opposite sign particle pairs to define the
reference for correlation function $C(Q)$\ shown in 
figure~\ref{fig:opalbe}(a).  However, this reference also contains
pairs coming from resonance decays and from weakly decaying particles.  
In addition, the correlation function has to be
normalised and suffers, at large four-momentum differences,
from long-range correlations due to energy and momentum conservation.  
Therefore OPAL use a large sample of Monte Carlo to define a second
correlation function $C'(Q)=C^{data}/C^{MC}$, 
shown in figure~\ref{fig:opalbe}(b).
This correlation function is more reliant on the Monte Carlo but 
is almost normalised with a reduced
contamination from correlated unlike charge pairs.  DELPHI and L3 both
use event mixing techniques to define their references.  

OPAL and DELPHI parameterise the two-pion correlation function in two
dimensions, while L3 use three dimensions.  OPAL also study a
one-dimensional correlation function as a function of the angle between the
two-pion momentum difference, in the rest frame of the $\pi\pi$\
system, and the thrust direction.  OPAL also check for any dependence 
on the two-jet nature of the events.  

In all cases, a significant difference between the transverse,
$r_{t}$, and longitudinal, $r_{l}$, dimensions is observed, indicating
that the emitting source of pions has an elongated shape, with the
longitudinal dimension about 1.3 times larger than the transverse
dimension.  For
instance, the parameter values obtained by fitting an extended Goldhaber
parameterisation to the correlation function $C'(Q)$\ of OPAL are:

\begin{eqnarray*}
r_{l}       &  = & 0.935  \pm 0.013_{stat}  \pm 0.026_{syst}\ \rm{fm} \\
r_{t}       &  = & 0.720  \pm 0.009_{stat}  \pm 0.044_{syst}\ \rm{fm} \\
r_{l}/r_{t} &  = & 1.30   \pm 0.03_{stat}   \pm 0.12_{syst} \\
\end{eqnarray*}

\section{Other results}

OPAL measure BE correlations in $K^{\pm}K^{\pm}$\ pairs~\cite{opalkk},
which confirms an earlier DELPHI result~\cite{dkk}, and implies  that
there must also be such correlations in $K^{0}_{s}K^{0}_{s}$\  pairs,
making it unlikely that previously observed threshold enhancements can
be attributed entirely to $f_{0}$(980) production.  SLD study
correlations in rapidity between identified charged hadrons and use
the SLC electron beam polarisation to tag the quark hemisphere in each
event, allowing the first study of rapidities signed such that the
positive rapidity is along the quark direction, which provides new
insights into the fragmentation process~\cite{sld}.  OPAL perform a
multidimensional study of local multiplicity fluctuations and genuine
multi-particle correlations in terms of factorial moments and, for the
first time in $e^{+}e^{-}$\ annihilation, factorial cumulants, up to
fifth order~\cite{opalint}.  The Monte Carlo models JETSET 7.4 and
HERWIG 5.9 are found to reproduce the trends but underestimate the
magnitudes.

\section{Summary and conclusions}

FD correlations have been clearly observed in
($\Lambda\Lambda$,$\bar{\Lambda}\bar{\Lambda}$) pairs by ALEPH using
two different techniques. A comparison of the source
size measured in systems of identical pions, kaons and $\Lambda$'s 
indicates that the source size decreases as the
mass of the emitted particles increases.  Several studies of
multidimensional BE correlations by OPAL, DELPHI and L3 confirm a
theoretical prediction that the emitting source of pions has an
elongated shape, with a longitudinal dimension  about 1.3 times
larger than the transverse dimension.  This indicates that models
based on the assumption of a spherical source are too simple.

\end{document}